\documentclass{article}
\usepackage{frascatiphys15lfc}
\usepackage{graphicx}
\begin{document}
\title{ 
A DATA-DRIVEN APPROACH TO 
PILE-UP  AT HIGH LUMINOSITY 
}
\author{
F Hautmann        \\
{\em 
 RAL, University of Oxford and University of Southampton} \\
}
\maketitle
\baselineskip=11.6pt
\begin{abstract}
We discuss recent results on pile-up based on a data-driven 
jet-mixing method. We illustrate 
prospects for 
experimental searches and 
precision  studies in 
high pile-up regimes at high-luminosity hadron 
colliders, showing 
how  the jet mixing approach  can be used,  
 also outside tracker  
acceptances,   to 
treat correlation observables and effects of hard jets from 
pile-up.\footnote{Talk given 
 at the Workshop  {\em LFC15}: Physics Prospects for  Linear and Other Future Colliders, 
ECT* Trento, 7-11 September 2015.} 
\end{abstract}
\baselineskip=14pt
%

Experiments at high-luminosity  
hadron colliders   face the 
challenges  of  very large  
pile-up, namely, a very  large number   of  overlaid 
hadron-hadron collisions per bunch crossing.  
At the Large Hadron Collider (LHC), 
for instance, 
in   data taken at Run I   
the pile-up is about 20 $ p p $ collisions on average,  
while it  reaches  the level of 
over 50 at Run II, and increases  
 for  higher-luminosity 
runs~\cite{ATLAS:2014cva,ATLAS:2014nea,TheATLAScollaboration:2013pia,Marshall:2014mza,CMS:2014ata,CMS:2013wea,Ghosh:2015raa,snowmass2013,Haas:2014talk,Hildreth:2014talk,Fartoukh:2014nga}. 

Current methods to deal with pile-up at the LHC employ 
precise vertex and track reconstruction, in regions  
covered by tracking detectors. More 
generally, they   
 rely on Monte Carlo 
simulations to 
 model  pile-up for data comparisons. This however 
 brings in a  model dependence 
which is particularly significant in 
regions where 
no detailed and precise measurements are available 
to constrain Monte Carlo  generators. 

Ref.\cite{Hautmann:2015yya} proposes 
a complementary approach to pile-up 
treatment, using data-driven methods 
rather than Monte Carlo modeling.  
The main purpose of this approach 
is to deal with potentially large contributions from jets with high transverse momenta, produced from pile-up events independent of the primary interaction vertex, in  a region 
where tracking devices are not available to identify pile-up jets. 
The goal is to treat not only inclusive 
observables but also correlations, and to rely 
 on  data recorded in high pile-up runs, rather than  requiring 
dedicated runs at low pile-up.  

\begin{figure}[htb]
  \begin{center}
\includegraphics[scale=0.4, trim=0cm 5.9cm 0cm 0cm, clip=true]{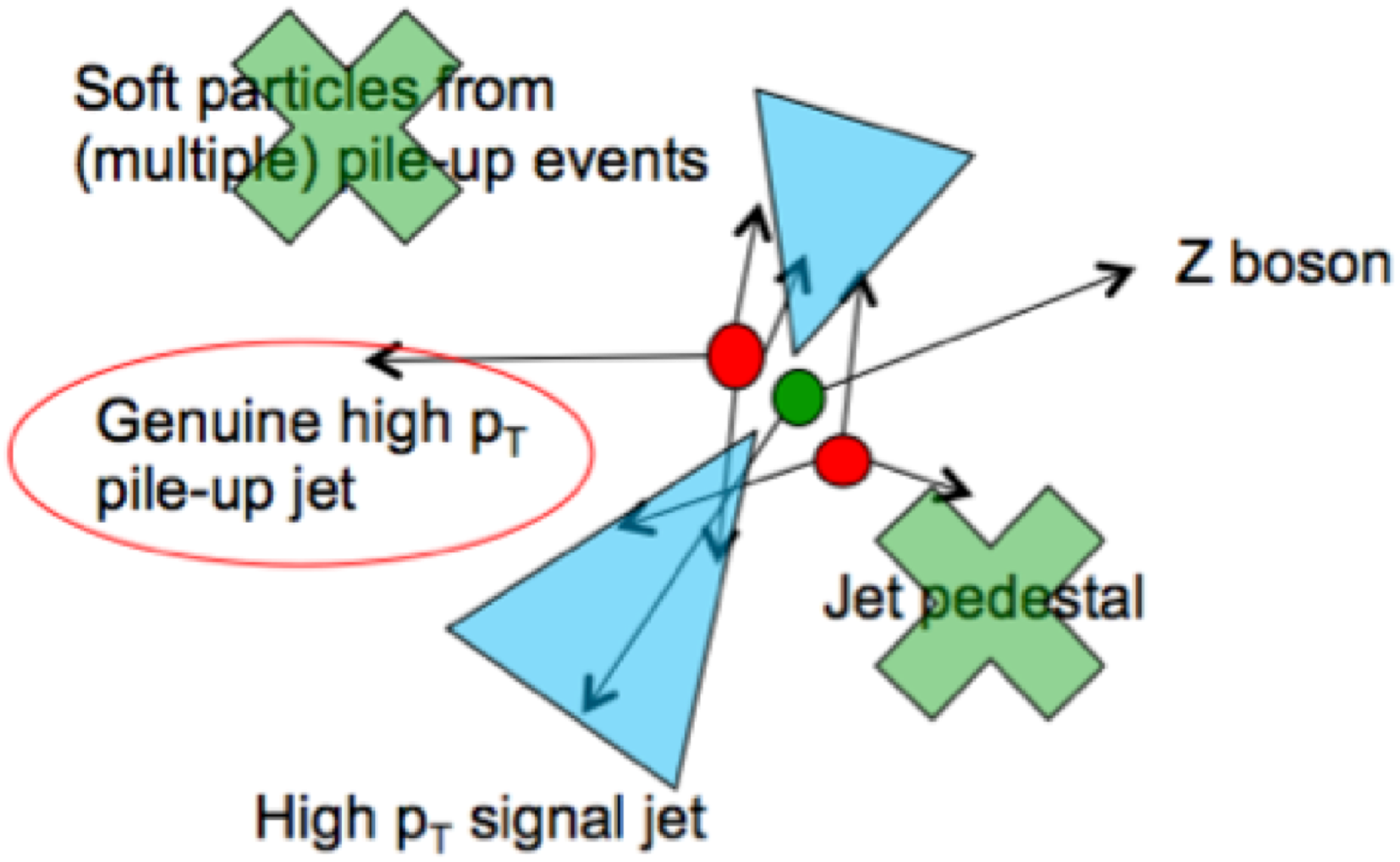}
  \caption{\it  Pile-up  contributions to the  reconstruction  of jets  in 
$Z$-boson + jet production. 
}
\label{fig:plp_fig_mid}
   \end{center}
\end{figure}

The basic idea of 
Ref.\cite{Hautmann:2015yya}  
can be illustrated using 
Drell-Yan lepton pair production associated with jets. This  can 
straightforwardly be  extended to a 
large variety of processes affected 
by pile-up. Fig.~\ref{fig:plp_fig_mid} 
shows a cartoon picture of 
different effects due to pile-up in 
$Z$-boson + jets production. 
One effect, denoted as jet pedestal, 
consists of additional pile-up 
particles in the jet cone, leading to  a 
bias in the jet transverse 
momentum. Another  is the 
overlapping of soft particles from 
pile-up, which are clustered into 
jets. A further effect is the 
misidentification of high transverse 
momentum jets produced from 
 independent pile-up events. 

Several methods exist 
to take the first two effects into 
account and correct for them. 
These include 
techniques based on the jet vertex 
fraction\cite{TheATLAScollaboration:2013pia} 
and charged hadron  subtraction\cite{CMS:2014ata,Kirschenmann:2014dla},  
the {\sc Puppi} method\cite{Bertolini:2014bba}, 
the SoftKiller method\cite{Cacciari:2014gra}, the 
jet cleansing method\cite{Krohn:2013lba}. 
These methods correct for transverse 
momenta of individual particles,  but 
not for any misidentification. 
The objective of the 
approach\cite{Hautmann:2015yya} 
is to analyze and treat the third 
effect, due to the mistagging of 
high transverse momentum 
pile-up jets. 

\begin{figure}[htb]
  \begin{center}
\includegraphics[scale=0.45, angle=90]{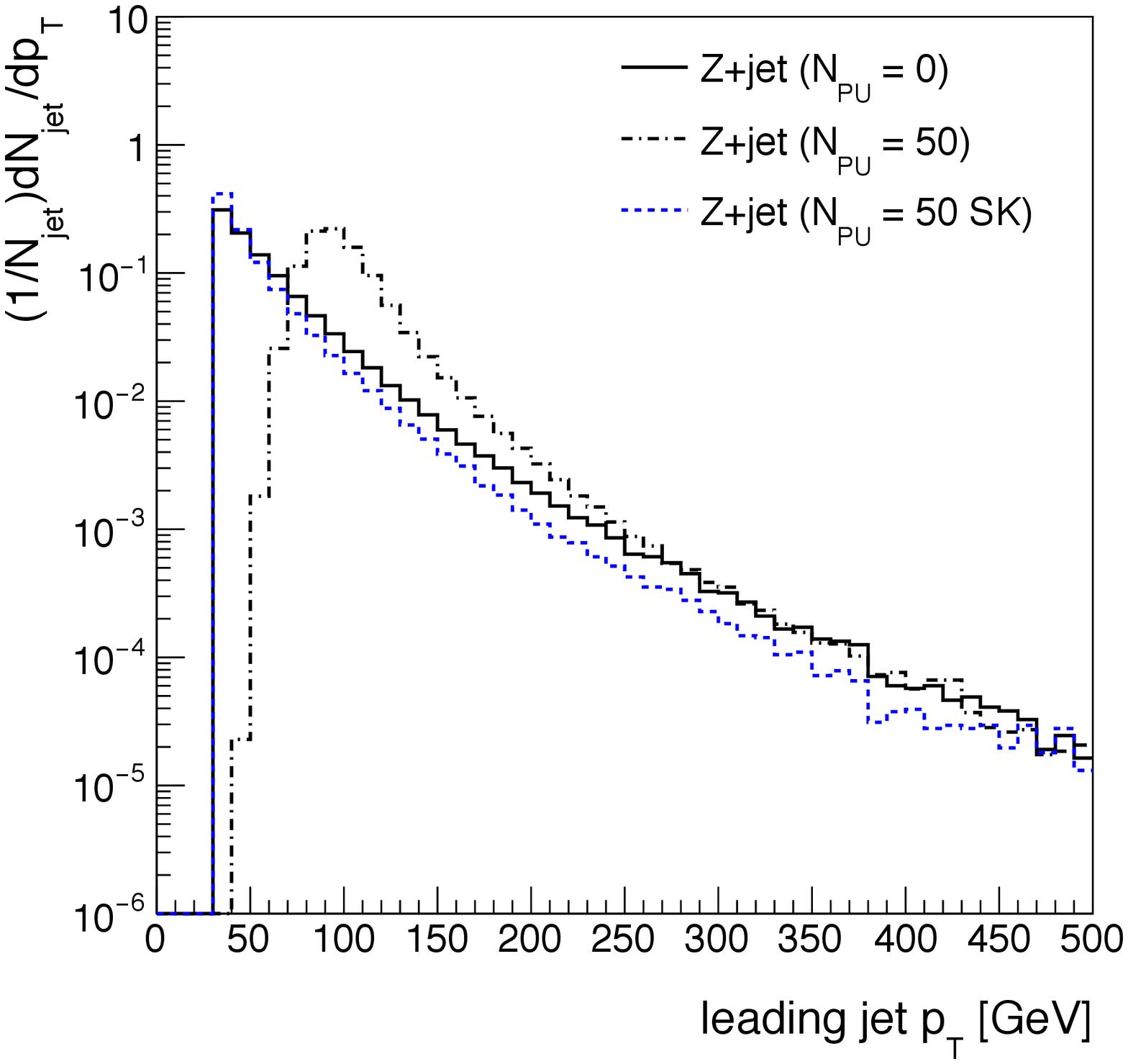}
\includegraphics[scale=0.45, trim=0cm 5.9cm 0cm 0cm, clip=true]{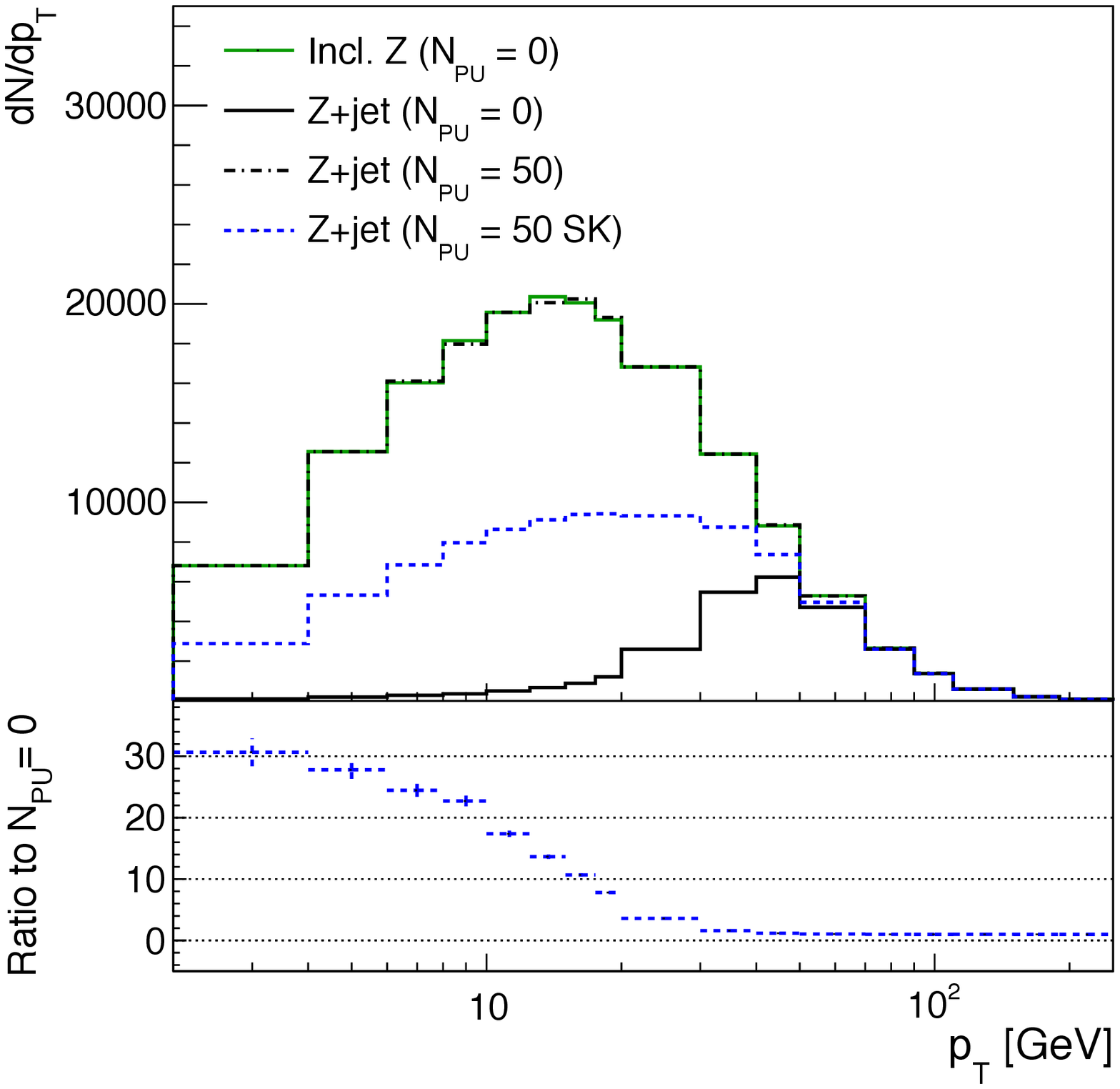}
  \caption{\it 
Effects  of pile-up  in 
$Z$-boson + jet production 
at the LHC: 
(top) the leading  jet $p_T$ spectrum; 
(bottom)  the  $Z$-boson $p_T$ spectrum\cite{Hautmann:2015yya}.}
\label{fig:plp_fig2}
   \end{center}
\end{figure}

Fig.~\ref{fig:plp_fig2}     illustrates 
the overall contributions of pile-up to 
$Z$-boson + jet correlation  variables.  
In the top plot is the leading jet $p_T$ 
spectrum, while 
in the bottom plot is the 
$Z$-boson $p_T$ spectrum. 
Event samples 
for $Z$-boson + jet production,  
with 
boson rapidity and invariant mass 
$ | \eta^{({\rm{boson}})}  |  <  2 $, 
60  GeV $  < m^{({\rm{boson}})} < $  120 GeV, and jet transverse 
momentum and rapidity 
$ p_T^{({\rm{jet}})} > 30 $ GeV, 
$ | \eta^{({\rm{jet}})}  |  <  4.5 $,
are generated, using the 
 anti-$k_T$ jet    algorithm\cite{Cacciari:2008gp} with 
distance parameter $R = 0.5$, 
by {{\sc Pythia}}  8\cite{Sjostrand:2007gs}  
with the 4C tune\cite{Corke:2010yf}  
for the different  scenarios of zero  pile-up and 
 $N_{\rm{PU}}$ additional  $pp$  collisions at $\sqrt{s} = 13$ TeV.  
The solid black curve is the signal, 
represented by the result in absence 
of any pile-up collision. The 
dot-dashed black curve is the 
result for $N_{\rm{PU}} = 50$ 
pile-up collisions.  The dashed blue 
curve is the result of applying the 
 method 
SoftKiller\cite{Cacciari:2014gra} to 
remove contributions of  soft 
 pile-up particles in the event.

We see from 
Fig.~\ref{fig:plp_fig2} 
that the effects of 
pile-up on 
$Z$-boson  + jet spectra 
are  large. Further we see 
that, while the leading 
jet   $p_T$ 
spectrum can be corrected well 
by the pile-up removal method 
SoftKiller,  the  
$Z$-boson $p_T$ spectrum is still 
affected by significant deviations 
from the signal even after applying SoftKiller. 
 Ref.\cite{Hautmann:2015yya}  interprets this as an effect  of 
mistagged pile-up jets, and devises 
an approach based on jet mixing to  treat it. 

\begin{figure}[htb]
  \begin{center}
\includegraphics[scale=0.45, trim=0cm 5.9cm 0cm 0cm, clip=true]{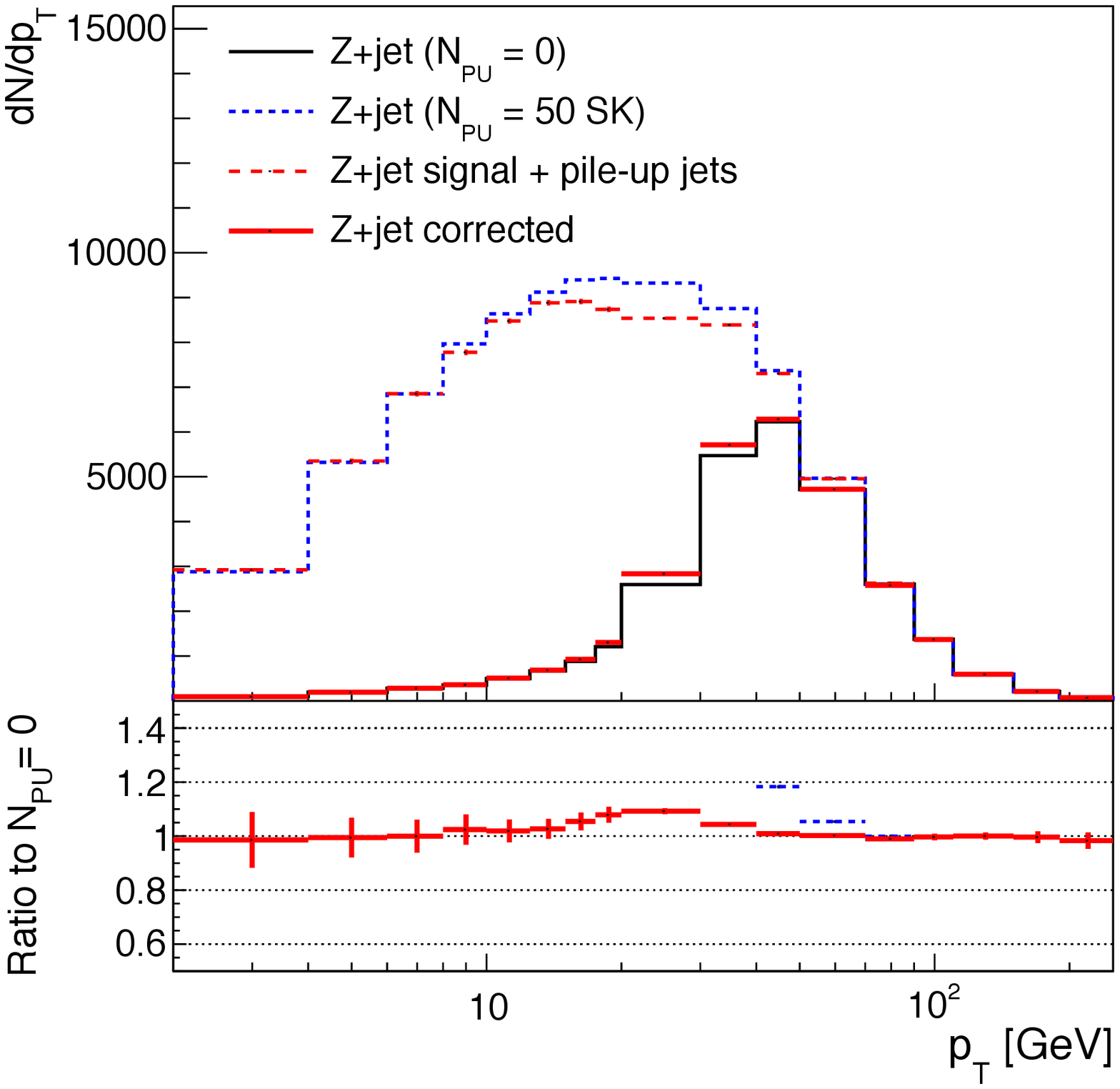}
\includegraphics[scale=0.45, trim=0cm 5.9cm 0cm 0cm, clip=true]{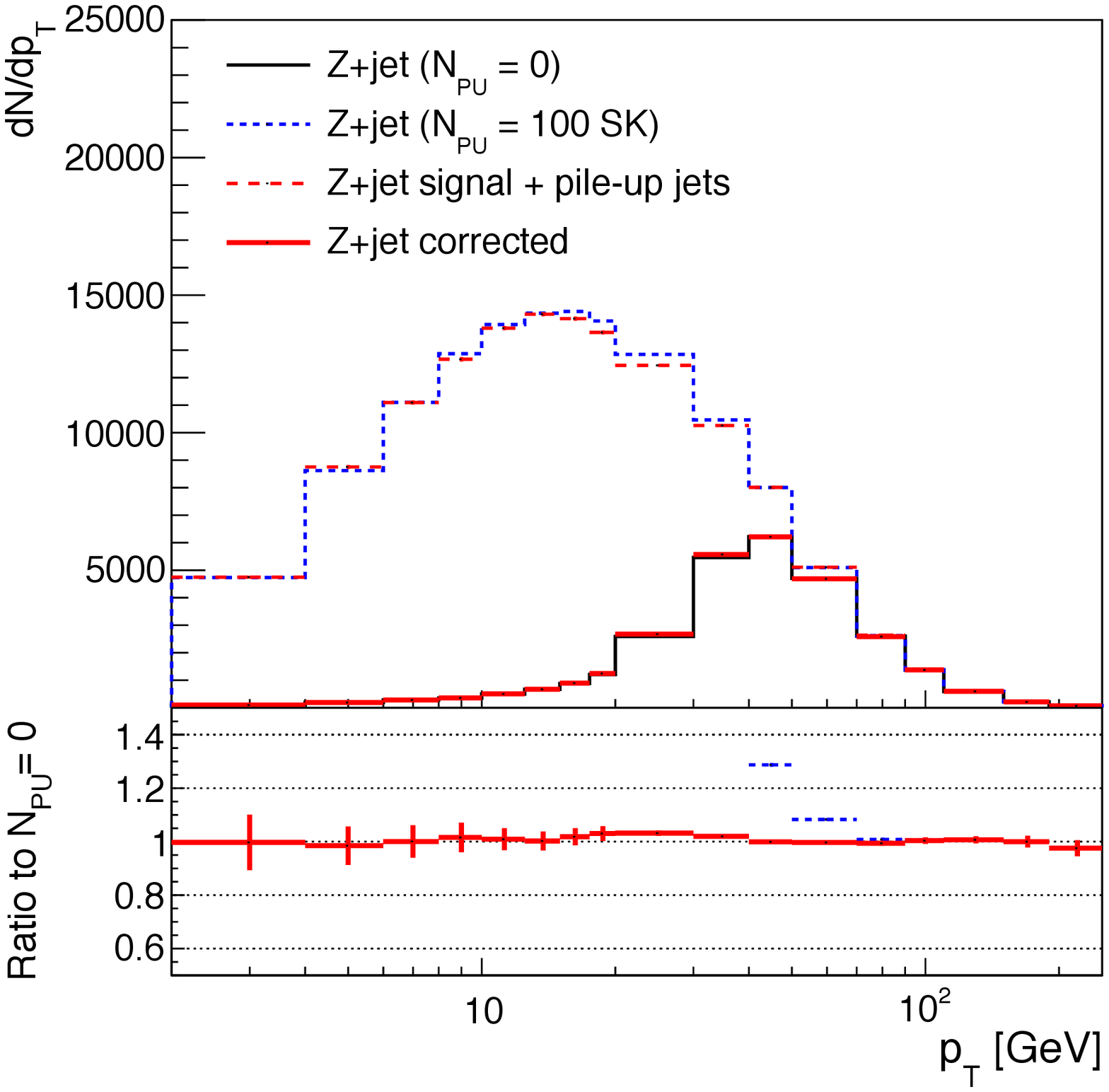}
  \caption{\it The  $Z$-boson $p_T$ spectrum  in $Z$ + jet production 
from the jet mixing 
method\cite{Hautmann:2015yya}: 
 (top)  
 $N_{\rm{PU}} = 50$;  (bottom)  $N_{\rm{PU}} = 100$. 
}
\label{fig:plp_fig3}
   \end{center}
\end{figure}

The jet mixing 
method\cite{Hautmann:2015yya} 
uses uncorrelated event 
samples to express the signal in the pile-up scenario  in terms 
of the signal without pile-up and a minimum bias sample of data at high pile-up. The results of this approach  
 are shown 
in 
Fig.~\ref{fig:plp_fig3},    
where 
samples containing   
$N_{\rm{PU}} $ minimum bias events 
are used for the mixing, in the cases 
$N_{\rm{PU}} = 50$ (top plot) and $N_{\rm{PU}} = 100$ (bottom plot). 
The solid black 
curve is the ``true" 
$Z$-boson plus jet signal.  
The dashed blue curve is the 
high pile-up, 
SoftKiller-corrected result   ($N_{\rm{PU}} = 50$ SK and $N_{\rm{PU}} = 100$ SK), representing 
the pseudodata in high pile-up. 
 The long-dashed red curve is the 
jet-mixed curve  obtained  from 
mixing the signal with the minimum 
bias sample.  The solid red curve is 
the final result, obtained 
by a simple ``unfolding", 
defined by multiplying the 
signal by  the ratio  of the 
pile-up (dashed blue) curve to the 
jet-mixed (long-dashed red) curve. 

We see from 
Fig.~\ref{fig:plp_fig3}    
that without the need to use    
Monte Carlo  events to simulate pile-up  the true signal 
is extracted nearly perfectly 
from the mixed sample.\footnote{Here we use a Monte Carlo 
to generate minimum bias  events  but under real running conditions this sample 
should be taken from real events recorded at high pile-up.} 
 This conclusion can be strengthened by 
checking\cite{Hautmann:2015yya}
that if the mixing is applied to a 
different starting distribution the unfolding still returns the true signal. 
Also, control checks on the jet resolution are carried out in 
Ref.\cite{Hautmann:2015yya},   
verifying that  features of the 
``true"  signal in  
the parton-jet 
 $p_T$  correlation and 
$ \Delta R$ distribution are 
well reproduced by the 
jet mixing. 
As Fig.~\ref{fig:plp_fig3}  indicates, the performance of the 
mixing technique, unlike the SoftKiller pile-up removal method,   improves 
with increasing $N_{\rm{PU}}$. 

In summary, the approach described 
in this article, while not addressing 
the question of a full detector simulation including pile-up, focuses on how to extract physics signals with the least dependence on pile-up simulation, and how to use real data, rather than Monte Carlo events, at physics object level. It can be applied 
to the high pile-up regime relevant for the LHC and for future 
 high-luminosity colliders, and does 
not require data-taking in special 
runs at low pile-up, so that there is no 
loss in luminosity.  It  implies good prospects both for  precision Standard Model studies 
at moderate scales 
affected by pile-up, 
e.g.~in 
Drell-Yan\cite{Dooling:2014kia,Angeles-Martinez:2015sea} and 
Higgs production\cite{Cipriano:2013ooa,VanHaevermaet:2014ela},  
and for searches for rare 
processes beyond Standard Model in high 
pile-up regimes.

\vskip 0.8 cm 

\noindent  {\bf Acknowledgments}. Many thanks to the organizers of 
the Workshop {\em LFC15}  for the invitation to a very interesting meeting.  
The results  presented in this article have been obtained in collaboration 
with H.~Jung and H.~Van Haevermaet.  Useful comments from M.~Dittmar, E.~Gallo, 
B.~Murray, P.~Van Mechelen and  M.~Wielers  
are  gratefully acknowledged. This work is supported in part by the 
DFG SFB 676 programme ``Particles, String and the Early Universe" at the 
University of Hamburg and DESY.

\end{document}